\documentclass[12pt]{article}
\usepackage{epsfig} 
\usepackage{amssymb}
\def\dash{\hbox{---}}
\def\abs#1{\left| #1\right|}
\def\pr#1{#1^\prime}

\def\beq{\begin{equation}}
\def\eeq{\end{equation}}

\def\beqn{\begin{eqnarray}}
\def\eeqn{\end{eqnarray}}
\relax

\newcommand\Ecm{E_{\rm cm}}

\catcode`\@=11

\@addtoreset{equation}{section}
\def\theequation{\thesection.\arabic{equation}}

\def\@normalsize{\@setsize\normalsize{15pt}\xiipt\@xiipt
\abovedisplayskip 14pt plus3pt minus3pt%
\belowdisplayskip \abovedisplayskip
\abovedisplayshortskip \z@ plus3pt%
\belowdisplayshortskip 7pt plus3.5pt minus0pt}

\def\small{\@setsize\small{13.6pt}\xipt\@xipt
\abovedisplayskip 13pt plus3pt minus3pt%
\belowdisplayskip \abovedisplayskip
\abovedisplayshortskip \z@ plus3pt%
\belowdisplayshortskip 7pt plus3.5pt minus0pt
\def\@listi{\parsep 4.5pt plus 2pt minus 1pt
     \itemsep \parsep
     \topsep 9pt plus 3pt minus 3pt}}
\@twosidetrue

\catcode`@=12

\newlength{\capwidth}
\catcode`\@=11

\def\section{\@startsection{section}{1}{\z@}{3.5ex plus 1ex minus
   .2ex}{2.3ex plus .2ex}{\large\bf}}

\def\thesection{\arabic{section}}

\def\appendix{\setcounter{section}{0}
 \def\thesection{Appendix \Alph{section}:}
 \def\theequation{\Alph{section}.\arabic{equation}}
  }

\catcode`\@=12

\def    \be             {\begin{equation}}
\def    \ee             {\end{equation}}
\def    \ba             {\begin{eqnarray}}
\def    \ea             {\end{eqnarray}}

\def    \=              {\;=\;}
\def    \frac           #1#2{{#1 \over #2}}

\def \as   {\alpha_{\rm s}}

\def \pt   {p_{\rm\scriptscriptstyle T}}

\def \mt   {\ifmmode m_{\rm T} \else $m_{\rm T}$ \fi}

\def \mur  {\mbox{$\mu_{\rm \scriptscriptstyle{R}}$}}
\def \muf  {\mbox{$\mu_{\rm \scriptscriptstyle{F}}$}}

\def \to   {\mbox{$\rightarrow$}}
\newcommand     \MSB            {\ifmmode {\overline{\rm MS}} \else
                                 $\overline{\rm MS}$  \fi}
\newcommand\ssrm{\scriptscriptstyle\rm}
\newcommand\nf{n_{\rm f}}
\newcommand\nlf{n_{\rm lf}}
\newcommand\Tf{T_{\ssrm F}}

\newcommand\Cf{C_{\ssrm F}}

\begin{document}
\begin{titlepage}
\nopagebreak
{\flushright{
        \begin{minipage}{4cm}
        CERN-TH/98-77  \hfill \\
        LPTHE-Orsay/98-11 \hfill \\
        IFUM/613-FT \hfill \\
        LNF-98/008(P) \hfill \\
        hep-ph/9803400\hfill \\
        \end{minipage}        }

}
\vfill
\begin{center}
{\LARGE 
{ \bf \sc
The $p_{\scriptscriptstyle \rm T}$ Spectrum in Heavy-Flavour
Hadroproduction }}
\vskip .5cm
{\bf Matteo Cacciari
\footnote{Work supported by an EC TMR Marie Curie fellowship under 
contract No. ERBFMBICT972126.}
}
\\
\vskip .1cm
{LPTHE, Universit\'e Paris-Sud, Orsay, France}
\vskip .5cm
{\bf Mario Greco}
\\
\vskip .1cm
{Dipartimento di Fisica E. Amaldi, Universit\`a di Roma Tre\\
and INFN, Laboratori Nazionali di Frascati, Italy}
\vskip .5cm
{\bf Paolo Nason\footnote{On leave of absence from INFN, Milan, Italy}}
\\
\vskip 0.1cm
{CERN, TH Division, Geneva, Switzerland} \\
%\vskip .4cm
\end{center}
\nopagebreak
\vfill
%\vskip 3cm
\begin{abstract}
  We consider the transverse-momentum distribution
  of heavy flavours in hadronic collisions. 
  We present a formalism in which large transverse-momentum
  logarithms are resummed at the next-to-leading level, and
  mass effects are included exactly up to order $\as^3$, so as
  to retain predictivity at both small and large
  transverse momenta.
  As an example, we apply our formalism to $b$ production at the Tevatron.
\end{abstract}
\vskip 1cm
CERN-TH/98-77 \hfill \\
March 20, 1998 \hfill
\vfill
\end{titlepage}

\section{Introduction}
Next-to-leading order calculations of heavy-flavour production have
been available for a long time \cite{Nason88,Nason89,Beenakker89,Beenakker91}.
These calculations exploit
the fact that the mass of the heavy quark acts as an infrared cutoff
on collinear singularities, and thus the cross section has a
power expansion in the strong coupling constant,
evaluated at a renormalization scale near the heavy-quark
mass $m$. This approach is appropriate when the heavy-quark
mass is the only relevant mass scale of the problem, and it is bound to fail
when the transverse momentum of the heavy quark is much larger
than its mass.
In fact, in this case, one cannot pinpoint a single characteristic
scale in the problem, since all momenta between $m$ and $\pt$
are equally involved. It turns out that,
whether we choose the renormalization scale $\mur$ and the factorization
scale $\muf$ of order
$m$ or of order $\pt$,
large logarithms of the ratio $\pt/m$ arise to all orders
in the perturbative expansion, and spoil its convergence.
The logarithmic
structure of the perturbative expansion for the inclusive
transverse-momentum distribution can be classified in terms of the
form $\as^2 (\as\log\pt/m)^k$ (which we call leading-logarithmic terms,
or LL), plus terms of the form $\as^3 (\as\log\pt/m)^k$ (next-to-leading
logarithmic terms, or NLL), and so on.

An early attempt to deal with this problem was described in
ref.~\cite{Nason89}, where an order-$\as^3$ calculation of the $\pt$
spectrum was presented.  There, in order to deal with the
logarithmically enhanced terms, the following approach was adopted: the
scales $\mur$ and $\muf$ were chosen of the order of $\pt$, and the
first neglected LL terms (that is to say, the terms of order
$\as^2(\as\log\pt/m)^2$) were estimated. Their value was
used as an error on the resulting cross section. This procedure, of
course, ends up giving large errors for very large $\pt$, but a
reasonable range of $\pt$ can be accessed in this way.

Alternatively, the whole tower of LL and NLL corrections can be
computed using the fragmentation-function formalism \cite{Cacciari94}.
This approach has however the drawback that it is essentially a ``massless''
formalism, in the sense that all contributions to the cross section
that are suppressed by powers of $m/\pt$ are not included.
Although these corrections must be small when $\log \pt/m$
is large, it is however never clear at what values of $\pt$ they can really
be neglected.

The purpose of this work is to construct a formalism in which
the fixed-order and the fragmentation-function approaches
are merged, in the following sense:
\begin{itemize}
\item
All terms of order $\as^2$ and $\as^3$ are included exactly,
including mass effects.
\item
All terms of order $\as^2\as^k \log^k \pt/m$
and $\as^3\as^k \log^k \pt/m$
are also included exactly.
\end{itemize}
To be more specific, let us write schematically the result of
the NLO calculation of the hadronic cross section as
\begin{equation}
  \frac{d\sigma}{d\pt^2}=A(m)\as^2+B(m)\as^3
  +{\cal O}(\as^4)\;.
\label{NLOsch}
\end{equation}
The explicit dependence upon the centre-of-mass energy $\Ecm$,
$\pt$ and $\mu$ is not indicated,
and $\as=\as(\mu)$.
The NLL resummed calculation is given by
\begin{eqnarray}
&&  \frac{d\sigma}{d\pt^2}=
        \as^2\sum_{i=0}^\infty a_i (\as\log \mu/m)^i
   +    \as^3\sum_{i=0}^\infty b_i (\as\log \mu/m)^i 
\nonumber \\&&
+ {\cal O}(\as^4(\as\log \mu/m)^i) + {\cal O}(\as^2 \times\mbox{PST})\;,
\label{NLLsch}
\end{eqnarray}
where PST stands for terms suppressed, in the large-$\pt$ limit,
by powers of $m/\pt$, irrespectively of further powers of logarithms
and of $\as$.
The coefficients $a_i$ and $b_i$ depend upon $\Ecm$, $\pt$ and $\mu$.
If $\mu \approx \pt$,
they do not contain large logarithms of the order of $\log \pt/m$.
The only large logarithms are the ones explicitly indicated.
Our approach combines the results of eqs.~(\ref{NLOsch}) and (\ref{NLLsch}),
giving
\begin{eqnarray}
&&  \frac{d\sigma}{d\pt^2}= A(m)\as^2+B(m)\as^3+
\nonumber \\&&
      \left(  \as^2\sum_{i=2}^\infty a_i (\as\log \mu/m)^i
   +    \as^3\sum_{i=1}^\infty b_i (\as\log \mu/m)^i  \right) \times G(m,\pt)
\nonumber \\&&
+ {\cal O}(\as^4(\as\log \mu/m)^i) + {\cal O}(\as^4\times\mbox{PST}) \;,
\label{FONLLsch}
\end{eqnarray}
where the function $G(m,\pt)$ is quite arbitrary, except that it must
approach 1 when $m/\pt\,\to\,0$, up to terms suppressed by powers of
$m/\pt$. Observe that the sums now start from $i=2$ and $i=1$,
respectively, in order to avoid double counting.  Thus, this formalism
contains all the information coming from the fixed-order NLO
calculation, and from the NLL resummed calculation.
The arbitrarity in the $G$ factor arises from the fact that we do not
know the structure of power-suppressed terms in the NLL resummed
calculation.

In the present work, we adopt the shortest path for obtaining the
correct answer, making use of an already existing computation of the
${\cal O}(\as^3)$ cross section ({\em fixed-order approach}, or FO)
and of already available computer codes to evaluate the resummed cross
section in the massless limit ({\em resummed approach}, or RS).

In order to carry out our program, it is important that both the RS
and the FO approaches are expressed in the same renormalization scheme.
The commonly used FO approach uses a renormalization and factorization
scheme in which the heavy flavour is treated as heavy.
Thus, if, for example, we are dealing with bottom production, we
use $\as$ of 4 light flavours as our running coupling constant,
and the appropriate structure functions should not include the bottom
quark in the evolution. The RS approach, on the other hand, also includes
the heavy flavour as an active, light degree of freedom.
This problem can be easily overcome by
a simple change of scheme in the FO calculation. 
Section \ref{sec:scheme} contains the details of this procedure.

Once this is done, the FO calculation matches exactly the terms
up to order $\as^3$ in the resummed approach, in the limit where
power-suppressed mass terms are negligible. In order to subtract from the RS 
result the fixed-order terms already present in the FO, we must provide
an approximation to the latter where only logarithmic mass terms are 
retained. We will call this ``massless limit'' FOM0.
In the simplified notation of eqs.~(\ref{NLOsch}) and (\ref{NLLsch})
we have
\begin{equation}
  A(m)=a_0+\mbox{PST}\,,\quad B(m)=a_1\log\mu/m+b_0+\mbox{PST}\;,
\end{equation}
and the FOM0 approximation is given by
\begin{equation}
  \frac{d\sigma}{d\pt^2}=a_0\as^2+\left(a_1\log\mu/m+b_0\right)\as^3
  +{\cal O}(\as^2 \times\mbox{PST})\;.
\label{FOM0sch}
\end{equation}

Our final result will be given by
\begin{equation}
  \label{eq:merge}
  \mbox{FONLL}=\mbox{FO}\;+\left( \mbox{RS}\; -\; \mbox{FOM0}\right)\;
\times G(m,\pt)\;.
\end{equation}
The notation FONLL stands for fixed-order plus next-to-leading logs.
Formula (\ref{eq:merge}) is our practical implementation
of eq.~(\ref{FONLLsch}).

An alternative approach to the problem has been given in
ref.~\cite{Olness97}, using an extension of a method developed for the
computation of heavy-flavour effects in deep-inelastic scattering
\cite{Aivazis94}. There, however, NLL terms are not fully included.  We
will comment on this approach in due time.

The paper is organized as follows. We first review in sect. \ref{sec:res}
the resummed approach, and in sect. \ref{sec:scheme}
we describe the procedure to adopt in order to translate the FO
result from a scheme with $\nf\,-\,1$ light flavours to a scheme with
$\nf$ light flavours. In sect. \ref{sec:FOM0} we give a few details
concerning the calculation of the massless limit of the FO calculation.
In sect. \ref{sec:matching} we check the matching between the FOM0 and
the RS calculation, by explicitly verifying that the difference RS-FOM0
is of order $\as^4$.
In sect. \ref{sec:powereff} we examine the size of power-suppressed
effects in order to understand at which value of $m/\pt$
the massless approach gives a sensible approximation to the
massive calculation. The function $G(m,\pt)$ will be chosen on the
basis of the considerations given in this section.
In sect. \ref{sec:LLmatching} we describe a simplified
version of our FONLL calculation, in which only LL resummation is adopted.
In sect. \ref{sec:dlscheme} we describe an alternative choice for
the factorization scheme for the fragmentation function.
In sect. \ref{sec:pheno} we describe our full result, for the case
of bottom production at the Tevatron.
Finally, in sect. \ref{sec:conclusions} we give our conclusions.

When computing specific cross sections, we will always use the
parton densities set CTEQ3M, which implements a
correct treatment of the bottom parton density.
Since this set uses a mass of 5~GeV for the bottom quark,
we will also employ the same value for consistency.

\section{Review of the resummation formalism}
\label{sec:res}
Essential ingredients for the resummation formalism
are the perturbative fragmentation
functions \cite{Mele91} for the parton $i$
to go into the heavy quark $h$, $D_i(x,\mu)$, where $i$ runs over all
the light partons (e.g. the light quarks and antiquarks, and the gluon)
plus the heavy quark and antiquark.
These fragmentation functions satisfy the standard Altarelli--Parisi
evolution equations, and their initial values at a scale $\mu_0$ of the order
of the heavy-quark mass $m$ are perturbatively calculable.
In the modified minimal subtraction scheme
($\overline{\rm{MS}}$) they are given by
\begin{eqnarray}
&&D_h(x,\mu_0) = \delta(1-x) + {{\alpha_s(\mu_0) \Cf}\over{2\pi}}\left[
{{1+x^2}\over{1-x}}\left(\log{{\mu_0^2}\over{m^2}} -2\log(1-x)
-1\right)\right]_+ \label{DQQ} \\ 
&&D_g(x,\mu_0) = {{\alpha_s(\mu_0) \Tf}\over{2\pi}}(x^2 + (1-x)^2)
\log{{\mu_0^2}\over{m^2}} \label{DgQ} \\
&&D_i(x,\mu_0) = 0 \label{DqQ}\,,\quad\mbox{for $i\ne g,h$}\,,
\end{eqnarray}
$h$ being the heavy quark and $g$ the gluon.
As usual, $\Cf = 4/3$ and $\Tf = 1/2$. The notation
$[f(x)]_+$ denotes the so-called $+$-distribution, whose integral
against any smooth function $g(x)$ is defined by the equation
\begin{equation}
 \int_0^1 g(x)\,[f(x)]_+\,dx = \int_0^1 \left(g(x)-g(1)\right)\, f(x) \,dx\;. 
\end{equation}

The perturbative fragmentation functions (PFF), evolved up to any 
scale $\mu\sim \pt$ via the Altarelli--Parisi equations, 
can be used to evaluate heavy-quark cross sections in the
large-transverse-momentum region by convoluting them
with short-distance cross sections for massless partons 
\cite{Aversa89,Aurenche86,Aurenche85}, subtracted in the  $\overline{\rm{MS}}$
scheme. The heavy quark is also treated as a massless active flavour, and
therefore also appears in the parton distribution functions of the colliding
hadrons and in the evolution of the strong coupling
constant\footnote{For this reason,
this resummation formalism is sometimes referred to (somewhat improperly)
as the ``massless scheme''.}.
The differential cross
section for the hadroproduction process
\begin{eqnarray*}
H_1(P_1) + H_2(P_2) \to h(P) + X \;,
\end{eqnarray*}
following the notations of \cite{Aversa89}, is given by
\begin{eqnarray}
{{d^2\sigma}\over{d\pt^2  dy}} &=& {1\over{S}} 
\sum_{ijk}\int^1_{1-V+VW} 
{{d z}\over{z^2}}
\int^{1-(1-V)/z}_{VW/z} {{d v}\over{1-v}} \int^1_{VW/zv} {{d w}\over w} 
\times\nonumber\\
&&\times F_{H_1}^i(x_1,\muf)F_{H_2}^j(x_2,\muf)
D_k(z,\muf)\times\label{resolved}
\\
&&\times \left[{1\over v}\left({{d\sigma^0(s,v)}\over{d v}}\right)_{ij\to k}
\delta(1-w) + {{\as^3(\mur)}\over{2\pi}}K_{ij\to
k}(s,v,w;\mur,\muf)\right] \, , \nonumber
\end{eqnarray}
having defined the hadron-level quantities
\beq
V = 1+{T\over S}\,,\qquad\qquad W = {-U\over{S+T}}\,,
\eeq
with $S=(P_1+P_2)^2$, $T=(P_1-P)^2$ and $U=(P_2-P)^2$.
We also define
\begin{equation}
  p_1=x_1\,P_1\,,\quad
  p_2=x_2\,P_2\,,\quad
  p = P/z
\end{equation}
and
\beq
s=(p_1+p_2)^2\,,\quad t=(p_1-p)^2\,,\quad u=(p_2-p)^2\,,\quad
v = 1+{t\over s}\,,\quad w = {-u\over{s+t}}\;.
\eeq
In terms of the momentum fractions $x_1$,
$x_2$ and $z$, it holds
\beq
s = x_1 x_2 S\,,\qquad x_1 = {{VW}\over{zvw}}\,,
\qquad x_2 = {{1-V}\over{z(1-v)}}\,.
\eeq
The $\sigma^0(s,v)$ terms represent the leading-order 
massless-parton to massless-parton scattering cross sections, while the
$K_{ij\to k}(s,v,w;\mur,\muf)$ represent the NLO corrections,
explicitly given in ref.~\cite{Aversa89}.

The implementation of the resummation procedure has been performed in 
ref.~\cite{Cacciari94} for $p\bar p$, in \cite{Cacciari96} for
$\gamma p$, and finally in \cite{Cacciari96a} for $\gamma\gamma$ 
collisions.
In all cases the results agree with the full massive ones
(refs. \cite{Nason89}, \cite{Ellis89} and \cite{Kraemer95}, respectively) in a
$\pt$ region from two to four times the mass of the heavy
quark. They have a smaller scale dependence
than the fixed-order calculations
at larger $\pt$, because the large logarithms originating from
gluon emission and gluon splitting are resummed by the evolution of the PFF.
In this region they are therefore more reliable
(see ref. \cite{Cacciari94} for a more complete discussion on this point).
Because of the intrinsically massless nature of the resummation procedure,
these results cannot be trusted when $\pt$ approaches $m$.

\section{The change of scheme}\label{sec:scheme}
When performing the full FO massive calculation one can,
according to ref.~\cite{Collins78}, conveniently define two
renormalization schemes that describe the same physics.
One is the usual \MSB\ scheme, in which all flavours are treated
on an equal footing. The other one, which we will call the decoupling
scheme, is similar to the \MSB\ scheme except for its treatment
of the divergences arising from heavy-flavour loops, which are
subtracted at zero external momenta. In the decoupling scheme,
for processes taking place at energies much below the
heavy-quark mass, we can forget about the heavy flavour altogether.
Thus, in QCD, one can define a standard \MSB\ scheme, as well as
decoupling schemes in which the top, or the top and the bottom,
or the top, the bottom and the charm are treated as heavy.
It is common to refer to the coupling
constant in the corresponding schemes as the 6-, 5-, 4- and 3-flavours
coupling constant.

We will now focus, for simplicity, on the case in which we have
$\nlf=\nf-1$ massless flavours and a heavy one of mass $m$.
The strong coupling constant in the decoupling scheme
and in the standard \MSB\ scheme
will be referred to as the $\nlf$ and $\nf$ flavours couplings respectively.
It turns out that the $\nlf$ and $\nf$  flavours couplings are identical at a
renormalization scale equal to the mass of the heavy quark:
\cite{Collins78}
\begin{equation}
  \label{eq:alphamatch}
\as^{(\nf)}(\mur)=\as^{(\nlf)}(\mur)+{\cal O}(\as^3)
       \qquad \mbox{for }\mur=m\,.  
\end{equation}
Thus, using the renormalization group equation we find
\begin{eqnarray}
\as^{(\nlf)}(\mur)&=&\as^{(\nlf)}(m) -
              b^{(\nlf)}_0\as^2 \log\frac{\mur^2}{m^2}
\label{eq:rengroup1}
\\
\as^{(\nf)}(\mur)&=&\as^{(\nf)}(m) -
              b^{(\nf)}_0\as^2 \log\frac{\mur^2}{m^2}\;\,
\label{eq:rengroup2}
\end{eqnarray}
where
\begin{equation}
  b_0^{(n)}=\frac{11 C_{\ssrm A}-4 n T_{\rm f}}{12\pi}\,.
\end{equation}
Observe that in the highest-order terms in eqs.~(\ref{eq:rengroup1})
and (\ref{eq:rengroup2}) we do specify neither the $\nf$ (or $\nlf$) label
nor the scale in $\as$, since the corresponding differences are of
higher order.
Taking the difference of eqs.~(\ref{eq:rengroup2}) and (\ref{eq:rengroup1})
and using eq.~(\ref{eq:alphamatch}) we arrive at
\begin{equation}
  \as^{(\nlf)}(\mur)=\as^{(\nf)}(\mur)
        -\frac{1}{3\pi}T_{\ssrm f} \log\frac{\mur^2}{m^2} \as^2
            +{\cal O}(\as^3)\,.
\end{equation}
Similarly structure functions for $\nlf$ and $\nf$ massless flavours
must match when $\muf=m$ \cite{Collins86}.
More specifically, they must satisfy the conditions
\begin{eqnarray}
F^{(\nf)}_j(x,m^2) &=& F^{(\nlf)}_j(x,m^2)\mbox{\ \ \ for\ }j\ne h
\nonumber \\
F^{(\nf)}_h(x,m^2)&=&0 \nonumber \\ \label{eq:pdfmatch}
F^{(\nf)}_{\bar{h}}(x,m^2)&=&0\,,
\end{eqnarray}
where $h$ stands for the heavy flavour.
It should be emphasized that this is a property of the \MSB\ 
subtraction scheme, and it is no longer true in other schemes.
Using the Altarelli--Parisi evolution
equations together with the matching conditions given in
eqs.~(\ref{eq:pdfmatch}),
one can easily find the appropriate relations between the parton 
densities with $\nlf$ and $\nf$ active flavours
for $\mu$ of the order of $m$.

We begin with the Altarelli--Parisi equations for the parton densities
with $\nf=\nlf+1$ flavours
\beq
\frac{\partial F^{(\nf)}_i(x,\mu) }{\partial \log \mu^2}=
\frac{\as^{(\nf)}(\mu)}{2\pi}\sum_j\int^1_x F^{(\nf)}_j(x/z,\mu)
P^{(\nf)}_{ij}(z)\frac{dz}{z}\;.
\eeq
For $\mu$ of the order of $m$, neglecting terms of order $\as^2$,
we get
\beq\label{eq:pdfpartons1}
F^{(\nf)}_i(x,\mu)-
F^{(\nf)}_i(x,m)=\frac{\as^{(\nf)}(m) {\displaystyle \log\frac{\mu^2}{m^2}}}
    {2\pi}
\sum_{j}\int^1_x
F^{(\nf)}_j(x/z,m) P^{(\nf)}_{ij}(z)\frac{dz}{z}.
\eeq
Observe that the heavy-quark density at $\mu=m$ vanishes because of
eqs.~(\ref{eq:pdfmatch}). We can then exclude it, 
putting $j\neq h,\bar{h}$ in the sum. For $i=h$ (or $i=\bar{h}$),
eqs.~(\ref{eq:pdfmatch}) and (\ref{eq:pdfpartons1}) yield
\beq\label{eq:pdfQ}
F^{(\nf)}_h(x,\mu)
=\frac{\as^{(\nf)}(m) {\displaystyle \log\frac{\mu^2}{m^2}}}{2\pi}
\sum_{j\neq h,\bar{h}}\int^1_x
F^{(\nf)}_j(x/z,m) P^{(\nf)}_{hj}(z)\frac{dz}{z},
\eeq
which shows that the heavy-quark density is of order $\as$.
An equation similar to eq.~(\ref{eq:pdfpartons1}) holds for $\nlf$ flavours:
\beq\label{eq:pdfpartons0}
F^{(\nlf)}_i(x,\mu)-
F^{(\nlf)}_i(x,m)=\frac{\as^{(\nlf)}(m) {\displaystyle \log\frac{\mu^2}{m^2}}}
{2\pi}
\sum_{j\neq h,\bar{h}}\int^1_x
F^{(\nlf)}_j(x/z,m) P^{(\nlf)}_{ij}(z)\frac{dz}{z}\,.
\eeq
Taking the difference between eqs.~(\ref{eq:pdfpartons1}) and 
(\ref{eq:pdfpartons0}), and using 
eqs.~(\ref{eq:pdfmatch}) and (\ref{eq:alphamatch}) we obtain,
for ${i\neq h,\bar{h}}$:
\beqn\nonumber &&
F^{(\nf)}_i(x,\mu)-
F^{(\nlf)}_i(x,m)=
\\
 \label{eq:pdfgdiff}
&&  \frac{\as^{(\nlf)}(m)\log{\displaystyle\frac{\mu^2}{m^2}}}
{2\pi}\sum_{j\neq h,\bar{h}}\int^1_x
 F^{(\nlf)}_j(x/z,m) \left[P^{(\nf)}_{ij}(z)
-P^{(\nlf)}_{ij}(z)\right]\frac{dz}{z}\;.
\eeqn
The only splitting function that depends explicitly upon the number
of light flavours is the gluon splitting function. We have
\beq
P^{(\nf)}_{gg}(z) - P^{(\nlf)}_{gg}(z)= -\frac{2\Tf}{3}\delta(1-z)\,,
\eeq
which applied to eq.~(\ref{eq:pdfgdiff}) gives
\beq
F^{(\nf)}_g(x,\mu)-
F^{(\nlf)}_g(x,m)= 
-\frac{\Tf\,\as^{(\nlf)}(m){\displaystyle\;\log\frac{\mu^2}{m^2}}}{3\pi}
F^{(\nlf)}_g(x,m)\;.
\eeq
The final result is then
\beqn
F^{(\nf)}_{h(\bar h)} &=&{\cal O}(\as)         \nonumber\\
F^{(\nf)}_j&=&F^{(\nlf)}_j+{\cal O}(\as^2)\mbox{\ \ \ for\ }
j\ne h(\bar{h}),\; j\neq g \nonumber \\ \label{eq:pdfhqeffect}
F^{(\nf)}_g&=&F^{(\nlf)}_g\left[1-\frac{\as 
\Tf}{3\pi}\log\frac{\mu^2}{m^2}\right]+{\cal O}(\as^2).
\eeqn
These equations are easily understood in the following way.
Since at $\mu=m$ the heavy-quark density is zero, it must be only of 
order $\as$ if $\mu$ is near but not equal to $m$.
The gluon density is affected at order $\as$ by the presence of the
heavy flavour. For $\mu>m$, there is in fact some more room for the
gluons to go into sea heavy flavours, and therefore the gluon
density is slightly diminished.
Light flavours do not couple directly to
the heavy one. They feel its effect only because of the diminished
gluon density, since there are less gluons to go into light flavours.
But the variation of the gluon density is itself of order $\as$,
and the probability for a gluon to go into a light fermion is also of 
order $\as$, so that the net effect is of order $\as^2$.

Now we have all we need to translate the heavy-flavour cross section
formulae from the decoupling scheme of ref.~\cite{Collins78}
to the the standard \MSB scheme.
First of all, the effect of the change of scheme on terms of order
$\as^3$ is of order $\as^4$, and can thus be neglected.
On the other hand, the Born term, of order $\as^2$,
will give rise to corrections of order $\as^3$.
In the case of the $q\bar{q}$ annihilation Born term, the quark densities
are unaffected at order $\as$, and thus the only correction arises
from the coupling constant. We have
\begin{equation}
 \sigma^{(0)}_{q\bar{q}} = \frac{\left[\as^{(\nlf)}(\mur)\right]^2}{m^2}
 f_{q\bar{q}}
=\frac{\left[\as^{(\nf)}(\mur)\right]^2}{m^2} f_{q\bar{q}}\left(1
 -\as \frac{2\Tf}{3\pi}\log\frac{\mur^2}{m^2}\right)+{\cal O}(\as^4)\;.
\end{equation}
In the case of the $gg$ fusion, instead, we obtain symbolically
\begin{displaymath}
\left[F^{(\nlf)}_g(\muf)\right]^2 \sigma^{(0)}_{gg}
 = \left[F^{(\nlf)}_g(\muf)\right]^2
\frac{\left[\as^{(\nlf)}(\mur)\right]^2}{m^2}
\;f_{gg}
\end{displaymath}
\begin{equation} \label{eq:ggtransf}
\quad\quad
=\left[F^{(\nf)}_g(\muf)\right]^2
\frac{\left[\as^{(\nf)}(\mur)\right]^2}{m^2}f_{gg}
\left(1
-\as \frac{2\Tf}{3\pi}\log\frac{\mur^2}{\muf^2}\right)\;.
\end{equation}

We thus summarize the results of the present section.
In order to compute fixed-order heavy-quark cross sections at the
${\cal O}(\as^3)$ level, using the standard \MSB\ scheme with $\nf$
flavours for both the coupling and structure functions, the modifications
one needs to apply to the partonic cross sections
of refs.~\cite{Nason88,Nason89}
are the following:
\begin{itemize}
\item
Add a term $-\as \frac{2\Tf}{3\pi}\log\frac{\mur^2}{m^2}\;
\sigma^{(0)}_{q\bar{q}}$ to the $q\bar{q}$ channel cross section.
\item
Add a term $-\as \frac{2\Tf}{3\pi}\log\frac{\mur^2}{\muf^2}\;
\sigma^{(0)}_{gg}$ to the $gg$ channel cross section.
\end{itemize}
Observe that, in case one uses $\muf=\mur$, there are no corrections
to the gluon channel. Furthermore, for any reasonable ranges
of scales and masses, these corrections are not large.

In the following, we will always refer to the FO and FOM0 calculations
performed in this transformed scheme. We will thus always
assume that $\as$ and the parton densities $F_j$ 
refer to $\as^{(\nf)}$ and $F_j^{(\nf)}$.

\section{Massless limit of the fixed-order calculation}
\label{sec:FOM0}
\begin{figure}[t]
  \begin{center}
%    \leavevmode
    \epsfig{figure=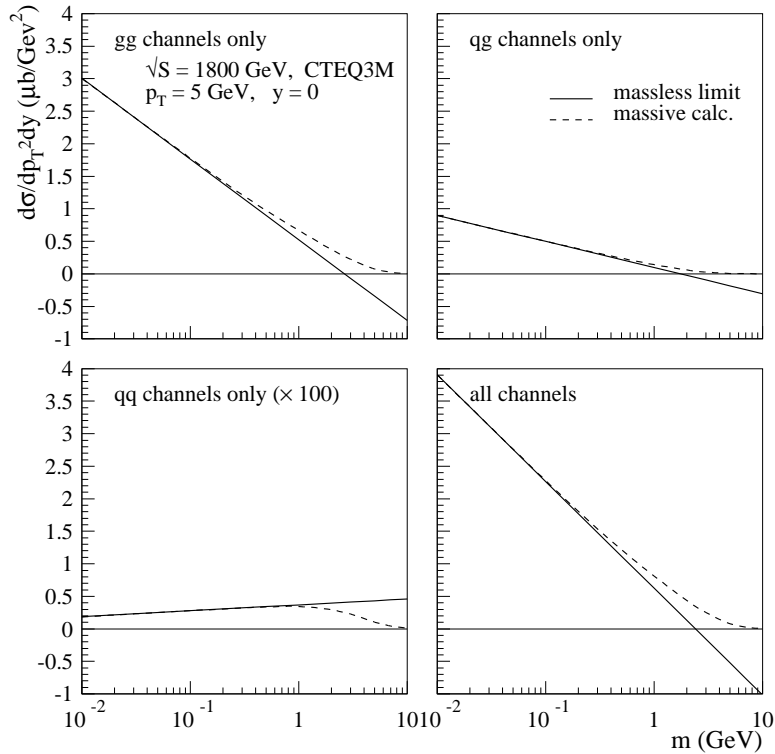,width=10cm}
\parbox{\capwidth}{
    \caption{\label{FOM0pt5} \protect\small
       Comparison of the FO and FOM0 differential 
       cross sections as a function of the logarithm of the mass,
       at $\pt=5$ GeV.}
}
  \end{center}
%    \label{FOM0pt5}
\end{figure}

\begin{figure}[t]
  \begin{center}
%    \leavevmode
    \epsfig{figure=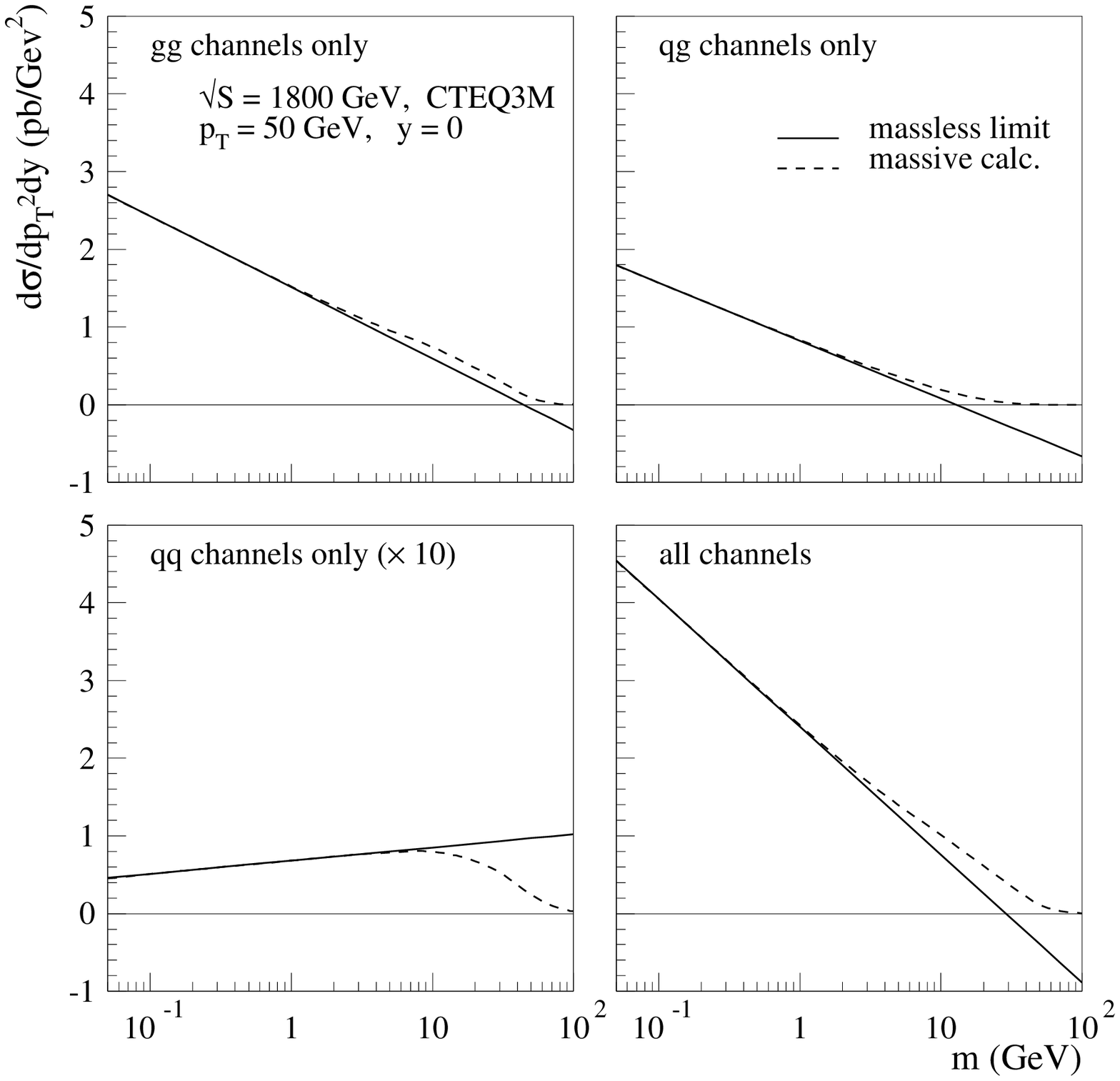,width=10cm}
\parbox{\capwidth}{
    \caption{\label{FOM0pt50} \protect\small
       Comparison of the FO and FOM0 differential cross 
       sections as a function of the logarithm of the mass, at $\pt=50$ GeV.}
}
  \end{center}
%    \label{FOM0pt50}
\end{figure}

The massless limit of the fixed-order formulae
(in the sense of eq.~(\ref{FOM0sch})) is obtained via algebraic methods
from the results of ref.~\cite{Nason89}. As pointed out there,
the limiting procedure is non-trivial. In fact, the partonic cross
sections at order $\as^3$ contain distributions,
such as delta functions or principal value singularities.
When taking the massless
limit, new contributions to these distributions arise.
We will not report here the rather boring details
of the algebraic procedure we followed. We would like to remark, however,
that its correctness can be easily checked in the following
way. We compute a heavy-flavour differential cross section at a fixed
$\pt$, rapidity, and centre-of-mass energy. We choose the renormalization
and factorization scales equal to $\pt$. Under these conditions,
the mass dependence of the result is confined to the partonic cross
sections. In the massless limit approximations, the only remnants
of mass dependence are in logarithms of the mass in the ${\cal O}(\as^3)$
terms. Thus, if we plot the FOM0 cross section versus the logarithm of
the mass, we get a straight line. On the other hand, if we plot the full FO
cross section versus the logarithm of the mass,
it should approach the FOM0 result in the limit of small masses.
This is in fact what we observe,
as can be seen in figs.~\ref{FOM0pt5} and \ref{FOM0pt50}.

From the figures, it is quite apparent that the massless limit,
as well as its implementation for the calculation of hadronic
cross sections, was carried out correctly.
There are also some important observations to make. The FOM0
cross section
is smaller than the massive calculation. On general grounds, one would 
instead expect the massive calculation to be smaller,
because of the reduction of phase space due to the presence of
masses. Moreover, at a given $\pt$, larger amounts of incoming
parton momenta are needed if the mass terms are kept,
which also reduces the cross section.
These arguments are in fact correct:
they should be applied to the absolute value of the cross sections
rather than to their signed value.
The value of the massive cross
section does become smaller and smaller as we approach the threshold.
It happens, however, that the massless limit cross section,
which has constant slope, changes sign at some point,
and then keeps growing in absolute value as we approach the threshold.

Notice that the FOM0 approximation is quite inaccurate even at
relatively small values of $m/\pt$. For example, from both
figs.~\ref{FOM0pt5} and \ref{FOM0pt50} we notice that already at
$m/\pt\approx 1/3$, the FO cross section is twice as large as the FOM0
result. In ref.~\cite{Cacciari94} it was observed that for bottom
production at the Tevatron, at $\pt \gtrsim 10\dash 15\,$GeV,
the resummed cross
section agrees quite well with the FO one (although no particular
reliability was expected for the resummed approach in that region).
Following the results displayed in this section, we must conclude that
such an agreement was accidental.

As a concluding remark for this section, we wish to point out that, in
principle, the FOM0 result could also be obtained by numerical
methods. It would suffice, for any given $\pt$, to evaluate the FO
cross section for a fictitious mass much smaller than $\pt$ and then
extrapolate linearly in the logarithm of the mass until the real $m$,
as can be seen from the aforementioned plots.  Needless to say, such a
method would however be extremely cumbersome, slower and less accurate
than the analytical one we have followed.

\section{Matching}\label{sec:matching}
We now examine the matching between the resummed approach and the FOM0
calculation.

There are ingredients in the resummed approach that are not explicitly
present in the FOM0 calculation. These are the fragmentation functions
for final-state partons to go into the heavy quark, and the parton density
for finding a heavy quark inside the hadron.
We take for simplicity
$\mur=\muf=\abs{\pt}$, and denote the common value with $\mu$.
The fragmentation function for any parton
to go into a heavy quark has a power expansion in terms
of the coupling constant evaluated at the scale $\mu$, and of
logarithms of $\mu/m$:
\begin{equation}
  \label{eq:Dseries}
  D_j(x,\mu,m)=\sum_{k=0}^\infty \sum_{l=0}^k
         d_j^{(k,l)}(x,\mu,m)\,\log^l\frac{\mu}{m}\,\as^k(\mu)\,,
\end{equation}
that can be obtained by solving the evolution equation for the fragmentation
function at the NLL level, with the initial conditions
(\ref{DQQ})--(\ref{DqQ}).
Similarly, the parton density for finding
the heavy flavour in the hadron can be expanded in the form
\begin{equation}
  \label{eq:Fseries}
  F_h(x,\mu,m)=\sum_{k=0}^\infty
 \sum_{l=0}^k f^{(k,l)}(x,F_l(\mu))\log^l\frac{\mu}{m}\,\as^k(\mu)\;.
\end{equation}
With $F_l(\mu)$ in the argument of the coefficients, we mean that the 
coefficients have a complicated {\em functional} dependence upon the parton
densities evaluated at the scale $\mu$.
The existence of formal expansions of the form~(\ref{eq:Dseries})
and (\ref{eq:Fseries}) can be easily proved, by writing the Altarelli--Parisi
equations in integral form, and then solving them iteratively.
We present a more detailed argument in Appendix A.

Once eqs.~(\ref{eq:Dseries}) and (\ref{eq:Fseries}) are formally
substituted in the RS cross section formula,
this formula itself becomes a power expansion of the form
of eq.~(\ref{NLLsch}), with the coefficients that depend
(functionally) upon the structure functions for light partons,
in the $\nf$-flavours scheme,
evaluated at the scale $\mu$. The FOM0 calculation has an expansion
of the same form (truncated to order $\as^3$) with coefficients
that are also dependent upon the same light-parton structure functions
\footnote{We observe that this property of the FOM0
calculation is only valid in the modified scheme described in section
\ref{sec:scheme}. If we had used the standard scheme for the fixed-order
calculation, the structure functions and the coupling $\as$
appearing there would be those with $\nf-1$ flavours.}
evaluated at the scale $\mu$.
Thus, because of the next-to-leading logarithmic accuracy of the
resummed cross section, the terms up to the order $\as^3$
will match exactly with the FOM0 calculation.

This property of the resummed calculation is built into the resummation
formalism of ref.~\cite{Cacciari94}. In this work, we have explicitly
verified this matching by comparing the numerical value
of the FOM0 and RS approaches in the limit $\as\,\to\,0$.

\begin{figure}[htb!]
  \begin{center}
%    \leavevmode
    \epsfig{file=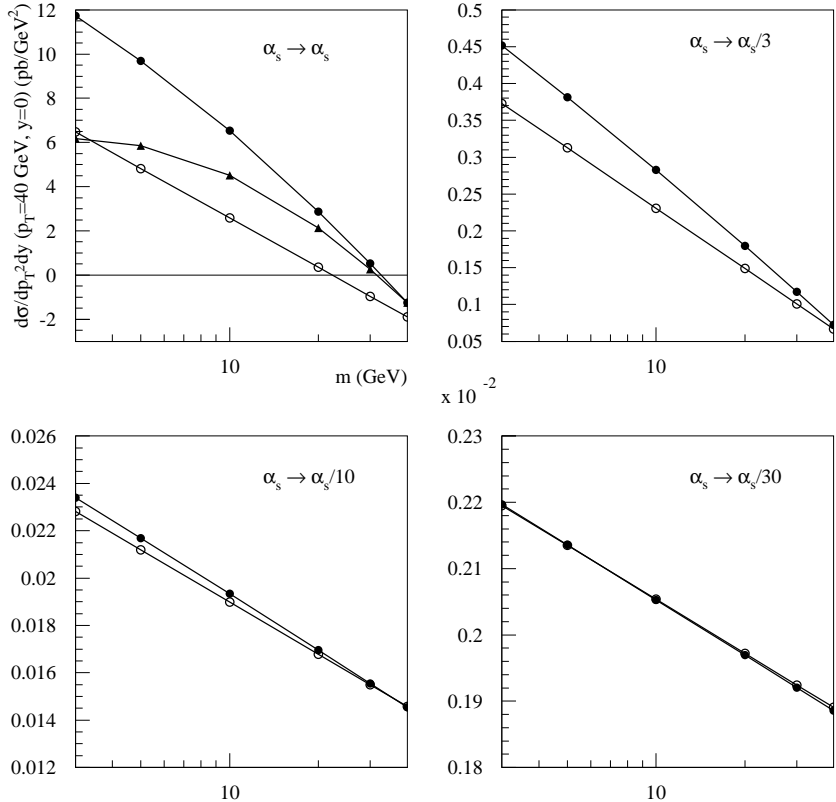,width=11cm,clip}
\parbox{\capwidth}{
    \caption{\label{fig:small_as} \protect\small Matching of the FOM0 and RS
     computations for small $\as$, for $\pt=40\;$GeV and $y=0$.
     The colliding hadrons are $p\bar{p}$ at $\sqrt{S}=1.8\;$TeV.
     The triangles represent the RS computation, the full circles the RSA,
     and the empty circles the FOM0.}
}
%    \label{fig:small_as}
  \end{center}
\end{figure}

In order to do this, we need to vary the coupling
constant towards very small values, in order to check if
the difference of the two approaches is indeed of order $\as^4$.
We thus need an explicit expression for the fragmentation functions
and for the heavy-flavour parton density in terms of $\as(\mu)$.
For our purpose, it is, however, sufficient to have an approximation
for these quantities that is valid at order $\as$.
These are easily obtained. For the heavy-flavour parton density we use
\beq\label{eq:pdfQapprox}
F_h(x,\mu)
=\frac{\as(\mu) \log\mu^2/m^2}{2\pi}
\int^1_x F_g(x/z,\mu) P_{hg}(z)\frac{dz}{z},
\eeq
which equals the corresponding expression in eq.~(\ref{eq:pdfQ}) up
to terms of higher order in $\as$.
%For the \MSB\ fragmentation functions, according to ref.~[\cite{MeleNason}],
%we can use
%\beqn \label{Dms1}
%D^{[1]}(x,\mu,m)&=&\delta(1-x)+\frac{\as(\mu) \CF}{2\pi}
%\left[\frac{1+x^2}{1-x}\left(\log\frac{\mu^2}{m^2}-2\log(1-x)-1\right)
%\right]_+ +{\cal O}(\as^2),
%\nonumber \\
%D_{[q\overline{q}\overline{H}]}(x,\mu,m) &=& {\cal O}(\as^2)
%\nonumber \\
%D_g(x,\mu,m) &=&
%\frac{\as(\mu) \Tf}{2\pi}(x^2+(1-x)^2)\log\frac{\mu^2}{m^2}+{\cal O}(\as^2).
%\eeqn
For the \MSB\ fragmentation functions
we instead simply use the initial conditions 
eqs.~(\ref{DQQ},\ref{DgQ}) evaluated at the large scale $\mu=\pt$
and without evolution. This is the first term in the expansion
of eq.~(\ref{eq:Dseries}).
We thus perform the RS computation with these approximate
parton densities and fragmentation functions; we call this approximation
RSA. In comparing RSA and FOM0 we vary $\as$ from its true value at the
given $\pt$ to 1/3, 1/10 and 1/30
of it. The results are collected in fig.~\ref{fig:small_as}.
The interpretation of the figure is quite obvious. The ${\cal O}(\as^2)$
result in the FOM0 calculation does not depend upon the mass of the
heavy quark, since it does not contain any logarithm. The slope of the
FOM0 result is only due to ${\cal O}(\as^3)$ terms,
as in figs.~\ref{FOM0pt5} and \ref{FOM0pt50}.
In fact,
we see that the relative slope of the result decreases roughly like $\as$,
and the relative difference between the RSA and the FOM0 result also
decreases as $\as^2$.
This demonstrates that the difference between the RSA and FOM0 results
is of order $\as^4$, as expected.
 
In the upper-left plot of fig.~\ref{fig:small_as}, the full RS
result (triangles) is also shown, evaluated with $\muf=\mur=\pt$.
We see that, if the mass is not too small,
the full RS result agrees reasonably well with the approximate RSA one,
the two being of course coincident when $m=\pt$, and therefore no evolution
whatsoever is taking place.

We observe that, at physical values of $\as$, and for
masses of the order of the transverse momentum, the RS cross section
differs significantly from the FOM0 one, in spite of the fact that
their difference is formally of order $\as^4$.
This is unfortunate, since it is precisely in this region
$(\pt\sim m)$ that we would like to see a smooth matching between the
FO result and the FONLL one, and this can only happen if the FOM0 and the RS
calculations cancel to a high extent.

\section{Power effects in the RS and FOM0 calculations}
\label{sec:powereff}
\begin{figure}[tb!]
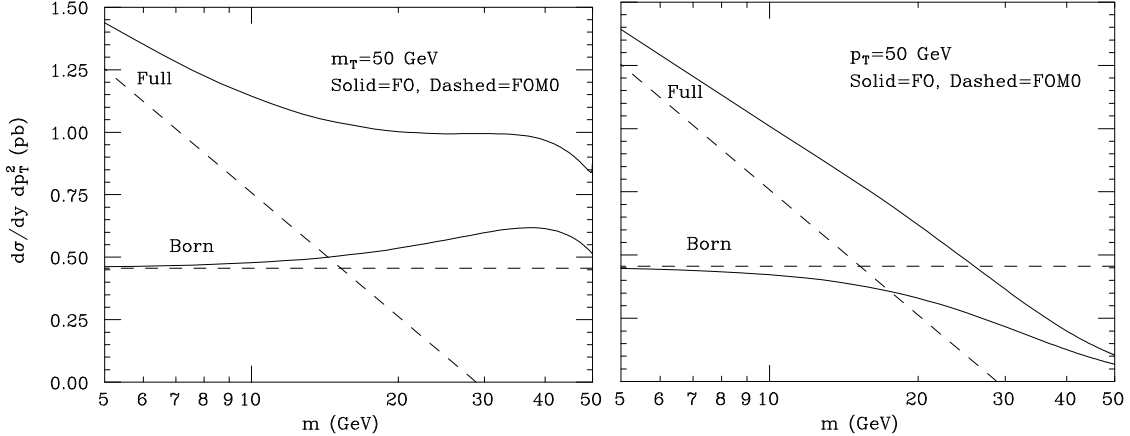

  \begin{center}
    \leavevmode
    \epsfig{figure=e_plot.eps,height=5.8cm}
    \epsfig{figure=p_plot.eps,height=5.8cm}
\parbox{\capwidth}{
\caption{\label{fig:pe_plots} \protect\small FO versus FOM0 at Born
              and full ${\cal O}(\as^3)$ level,
             plotted as a function of the mass and at fixed transverse mass
             (left figure) or fixed transverse momenta (right figure).}
}
%    \label{fig:pe_plots}
  \end{center}
\end{figure}

When comparing and matching the FO, FOM0 and RS approaches, there
is a lot of arbitrariness in the way mass effects are treated.
For example, we may decide to compare transverse-mass distributions
instead of transverse momenta. These are equal for the FOM0 and RS
calculations, but differ in the FO case.
The difference is shown in the two plots of fig.~\ref{fig:pe_plots},
where the cross section is plotted as a function of the mass,
keeping either $\mt$ or $\pt$ fixed.
If we focus upon the Born result, it is quite easy to understand
the differences in the two plots.
The massive result is closer to the massless one when the transverse
mass is kept fixed, i.e. in the left plot.
In fact, in this case, the differences come
only from mass effects in the matrix elements, and these effects are
not large. On the other hand, if the transverse momentum is kept fixed
(right plot),
as the mass grows the FO Born result requires more incoming momenta
from the structure functions, and this suppresses the cross section.
This mechanism is also at work, of course, in the NLO result.
However, in this case, the FOM0 approximation is bound to fail more radically,
because of its linear dependence in $\log \pt/m$: for large masses,
it changes sign, which is clearly unphysical. Of course,
it could be argued that the correct form of the logarithm in the massless
approximation should be something like  $\log (\pt^2+m^2)/m^2$,
but the point is that the only use of the FOM0 computation is to
subtract it from the available RS result, which only contains logarithms
of $\pt/m$. 
%In our phenomenological analysis we will therefore
%use the approach of matching the transverse mass of the FO
%approach to the transverse momentum of the RS and FOM0 approaches.

We will therefore proceed as follows.
For a given transverse momentum $\pt$, the FO cross section is evaluated
and combined, using eq.~(\ref{eq:merge}), with the FOM0 and RS results
evaluated instead at the corresponding $\mt = \sqrt{\pt^2 + m^2}$ value.
In this way the three calculations are performed at the same $\mt$.
Moreover, a central choice for the renormalization and factorization scales
will be $\mur=\muf=\sqrt{\pt^2 + m^2}$, so that they coincide
in the three calculations.

Having done so, one has further freedom in the choice of the
function $G(m,\pt)$, the factor multiplying
the RS$\,-\,$FOM0 term in eq.~(\ref{eq:merge}).
$G(m,\pt)$ should approach 1 at large $\pt$.
It could in fact be chosen equal to 1. However, as we showed
in sect.~\ref{sec:matching},
the difference RS$\,-\,$FOM0, although of order $\as^4$, is abnormally
large. We interpret this as a consequence of the fact that, when
$\pt$ is near $m$, the massless approximation becomes completely meaningless.
This is also visible from the left plot of
fig. \ref{fig:pe_plots}, where we can see that the FOM0 result starts
to deviate significantly from the full FO one when the mass becomes
larger than about one
fifth of $\mt$.
For a physical $b$ quark, with a mass of 5 GeV, this
means that we should suppress the RS$\,-\,$FOM0 term for $\pt$
smaller than about 20--25 GeV.
This seemingly large value
is in fact not so difficult to justify. The dominant logarithmic
effects come in fact from the flavour-excitation and gluon-splitting
graphs.
In order for a gluon-splitting phenomenon to be a truly collinear
process, and thus dominant over power-suppressed terms,
the incoming gluon must carry a transverse mass
that should be more than four times the mass of the heavy quark,
in such a way that the produced heavy quark and antiquark are both
relativistic. Most of this momentum should end up in the quark
(the cases in which the momentum is evenly shared is suppressed
by the luminosity). A similar reasoning also applies to the 
flavour excitation case.

We thus choose $G(m,\pt)=\pt^2/(\pt^2+c^2m^2)$,
and our final formula becomes
\begin{equation}
  \label{eq:merge2}
  \mbox{FONLL}=\mbox{FO}\;+\; {{\pt^2}\over{\pt^2 + c^2 m^2}}
  [\mbox{RS}\;-\; \mbox{FOM0}]\;,
\end{equation}
with $c=5$, which suppresses the resummation correction RS$\,-\,$FOM0
for $\mt < 5 m$.

Before closing this section it is important to recall how all these
manipulations involving power-suppressed mass terms
only affect terms of order $\as^4$ or higher. Whatever theoretical uncertainty
might stem from them, we must live with it, as these terms have never
been computed.

\section{An LL implementation of the matching procedure}
\label{sec:LLmatching}
We will now focus on a first implementation of our formalism,
in which only leading logarithms are resummed. Our resummed
formula will have the following form
\beq \label{eq:merge0}
{\rm FOLL} = \mbox{FO}\;-\; \mbox{FOM0LL}\;+\; \mbox{RSLL}\;,
\eeq
where for simplicity we have chosen $G(m,\pt)=1$.
The meaning of the various terms is as follows:
FOLL is our fixed-order plus leading-log result, FOM0LL is the
leading logarithmic part of FOM0, and RSLL is the leading logarithmic part
of the RS result. Thus FOM0LL contains terms proportional to $\as^2$
and $\as^3 \log \mu/m$, but no terms proportional to $\as^3$ alone.
More specifically, in the notation of eq.~(\ref{FOM0sch}) we have
\begin{equation}
  \mbox{FOM0LL}=a_0\as^2+a_1\,\log\frac{\mu}{m}\,\as^3\;.
\label{FOM0LLsch}
\end{equation}
The FOM0LL contribution
precisely cancel the terms up to order $\as^3$ in RSLL.
Thus, the difference $\mbox{RSLL}-\mbox{FOM0LL}$ begins with terms
of order $\as^4\log^2 \mu/m$.
Therefore, we have an interpolating formula that reduces to the NLO result
for small $\pt$, and sums all the leading logarithms in the
large-$\pt$ limit.
\begin{figure}[t]
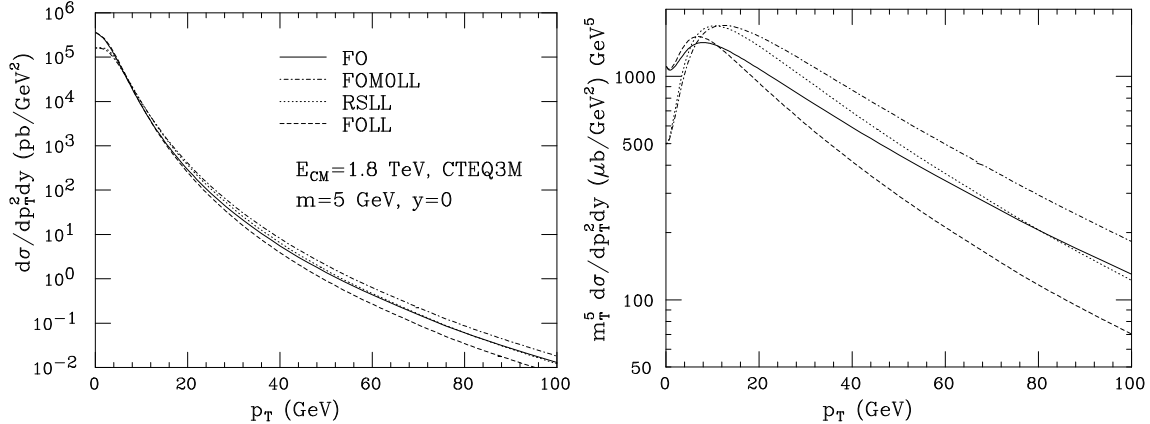

  \begin{center}
%    \leavevmode
    \epsfig{figure=ll.eps,width=7.5cm} \epsfig{figure=ll-mt.eps,width=7.5cm}
\parbox{\capwidth}{
    \caption{\label{ll} \protect\small
       Illustration of the FOLL result for the central values of
       the scales $\muf=\mur=\mt$.}
}
  \end{center}
\end{figure}
\begin{figure}[t]
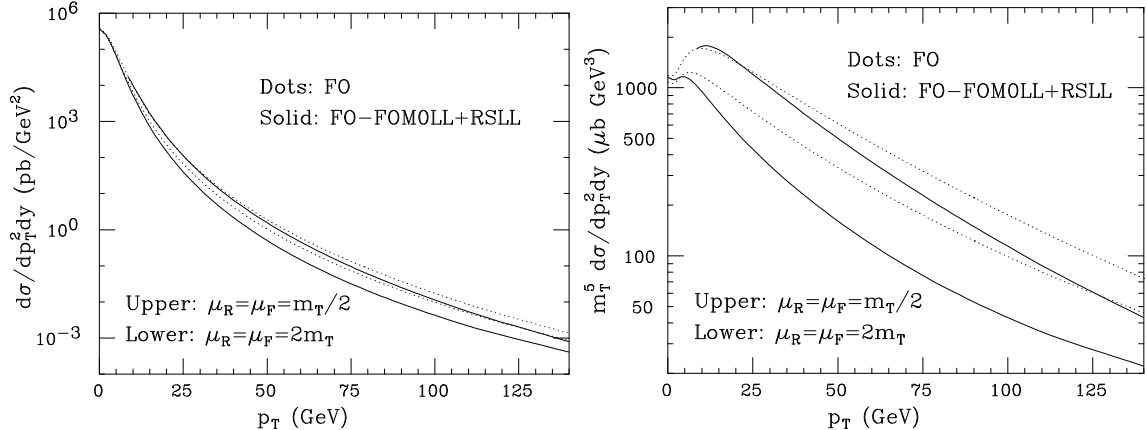

  \begin{center}
%    \leavevmode
    \epsfig{figure=band-ll-c0.eps,width=7.5cm}
    \epsfig{figure=band-ll-c0-mt5.eps,width=7.5cm}
\parbox{\capwidth}{
    \caption{\label{band-ll} \protect\small
       Illustration of the FOLL result for two scale choices.}
}
  \end{center}
\end{figure}
Our FOLL results are illustrated in figs.~\ref{ll} and \ref{band-ll}.
From fig.~\ref{ll}, which is obtained with a choice of scales
$\mur=\muf=\mt$, we see that the FOM0LL and the RSLL curves coincide at
$\pt=0$, corresponding to $\mu=\mt=m$. This is because their difference
is made up of terms of the form $\as^2(\as\log\mu/m)^k$ for $k\ge 2$,
and thus vanishes for $\mu=m$. They also depart relatively slowly from
each other, since there are at least two powers of logarithms
in their difference. As a consequence of this fact, the difference
between the FO and the FOLL curves vanishes at small $\pt$.
The FOLL curve undershoots instead the FO curve at larger $\pt$, in agreement
with what was estimated in ref.~\cite{Nason89}.

There is a basic difference between the FOLL approach and the FONLL
one. In the latter approach, it was mandatory to keep the same
scales $\mur$ and $\muf$ in all terms of the defining equation
(\ref{eq:merge}). In eq.~(\ref{eq:merge0}), on the other hand,
while it is important to maintain the same scale in the last two terms
(in order for their difference to be of order $\as^4$), this scale
could differ from the one used in the FO
term. In fact, the difference RSLL$\,-\,$FOM0LL is a sum of terms
proportional to $\as^2(\as\log\mu/m)^k$, and their scale variation
is thus of next-to-leading order (i.e. it contains an extra power of $\as$
or one less power of $\log\mu/m$). Thus, by a change of scale in the
RSLL$\,-\,$FOM0LL difference with respect to the FO scale,
one obtains an equivalent formula, up to the addition of
next-to-leading logarithmic terms, which are not fully specified in this
approximation.

This observation leads to the conclusion that no improvement in the
scale dependence can be observed if only LL resummation is
performed, or better, if an improvement is observed it must
be accidental. The reader may now wonder where the improvement lies
in this case of LL resummation, since undoubtedly we did resum
a tower of dominant terms.
One can easily convince oneself that the improvement lies in the
fact that, for a choice of scales near $\pt$, the large logarithms
in eq.~(\ref{eq:merge0}) are resummed, and thus one is justified
to consider only scale variations for scales of the order of
$\pt$. This point may sound confusing, since for heavy-flavour cross
sections at high transverse momenta the error is usually estimated
by varying the scale in a neighbourhood of $\pt$.
We remind the reader that this procedure is not obvious at all:
since we are facing a two-scale problem, one should vary the scale
between $m$ and $\pt$. The justification for using the upper range
$\mu \approx \pt$ only
was outlined in ref.~\cite{Nason89}. In that paper the scale
was varied in the neighbourhood of $\pt$, but subleading logarithms
were estimated, and included in the error. Since they turned out to be small,
they were no longer included in subsequent works.

Having said that, we do not expect a scale improvement in the
FOLL formula. The scale dependence of our FOLL result, displayed
in fig.~\ref{band-ll}, is indeed not smaller than the scale
dependence of the FO result.

It is now time to comment on the results of ref.~\cite{Olness97}.
There, the NLO fixed-order result is fully included in the
calculation, but the resummation is only performed with
leading-logarithmic level accuracy. The authors do find a
reduction in the scale dependence at moderate $\pt$,
while at large $\pt$ they find stronger scale dependence.
From the considerations given in the present section,
we conclude that, in fact, the scale compensation they observe
can only be accidental.

\section{Alternative choice of fragmentation scheme}
\label{sec:dlscheme}
As pointed out in sect.~\ref{sec:LLmatching}, the RS and FOM0 results
do not agree at the point $m=\pt$ (for the choice of scale $\mu=\pt$),
where we would like to see, instead,
a cancellation.  Furthermore, also the slopes of the two curves
are quite different at this point.
These two differences can easily be traced back
to two kinds of terms that arise in the RS calculation when
convoluting the NLO, \MSB\-subtracted massless kernels with the
structure and fragmentation functions.
The offset of RS with respect to FOM0 is due to spurious $\as^4$ terms, 
originating from the product of ${\cal O}(\as^3)$ terms in the kernel cross 
sections with the non-logarithmic ${\cal O}(\as)$ term in the $D_h$
fragmentation function initial condition of eq.~(\ref{DQQ}).
These terms do not contain logarithms of $\pt/m$, and thus remain
different from zero at $\pt=m$. They
are not correctly predicted by the RS calculation,
which is accurate up to terms proportional to $\as^4\log(\pt/m)$.
The difference in slope is, instead, due precisely to terms of this kind.
They are correctly given by the
RS formalism, but not included in the FOM0 result.

While nothing can be done to solve this second problem,
it is instead possible
to try to eliminate the first one. This can be done by switching from
the standard \MSB\ scheme to a new factorization scheme where the initial
condition for the heavy-quark to heavy-quark fragmentation function, 
eq.~(\ref{DQQ}), does not contain the non-logarithmic ${\cal O}(\as)$ term
and is therefore a delta-function for $\mu_0=m$.
We call this scheme the {\em Delta scheme} (DL).
The transformation to go from the \MSB\ scheme to the DL scheme
is described in detail in Appendix B.
\begin{figure}[htb]
\begin{center}
\epsfig{file=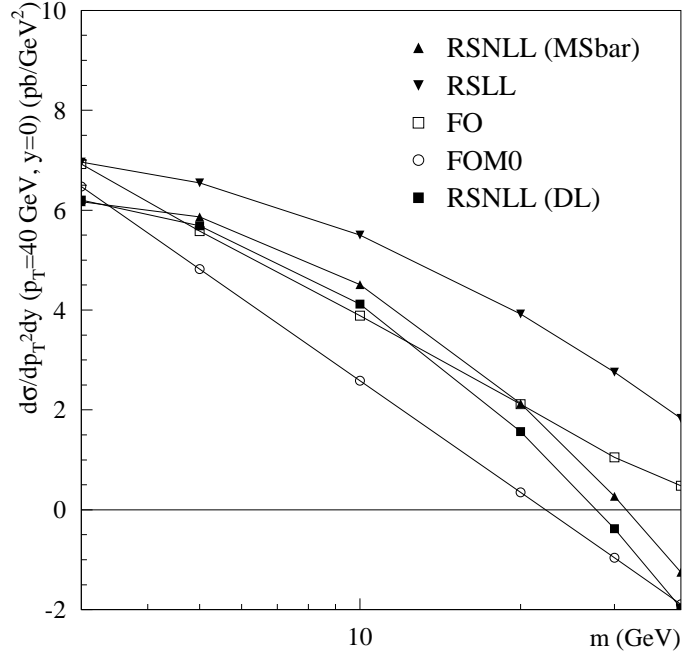,width=9cm}
\parbox{\capwidth}{
\caption{\label{fig:dl} \protect\small
Dependence of the cross section on the logarithm of the mass, at 
$\pt=40$ GeV. Shown are the full FO result (empty squares), the FOM0 massless
limit (empty circles), the LL resummed result (downward-facing triangles),
the RS result both in the \MSB\ (upward-facing triangles) and in the DL (full
squares) schemes.}
}
\end{center}
%\label{fig:dl}
\end{figure}
The effect of the scheme change can be appreciated in fig.~\ref{fig:dl}, 
where the cross section is again plotted as a function
of the logarithm of the mass. One
can clearly see how the DL result agrees with the FOM0 one at the
point $m=\pt$, thereby providing the desired cancellation.
However, the slope remains different.

For comparison, we also plot the RSLL result. This does not contain
the $\as^4\log(\pt/m)$ terms, and has therefore the same
slope as the FOM0 result.
Its offset with respect to the FOM0 result is instead due to
the non-logarithmic $\as^3$ term, which is present there.

\section{Some numerical NLL results}\label{sec:pheno}
%Final bands (try cteq4? at least verify b in cteq4 versus b in cteq3).
%Problem with fragmentation scale < m.
%Figures:
%1) central result for massive, DL, MS, LO
%2) LO con FO,
%3) DL con FO,
%4) MS con FO
%
%Variation of initial fragmentation scale (tutto da studiare)
%
%Osservazioni: c'e' un contributo positivo a pt intermedi?
%Restrizione della banda nel calcolo NL.
This section collects the final NLL results for the
heavy quark cross section, as obtained from combining the fixed-order massive
result with the resummed one.
\begin{figure}[t]
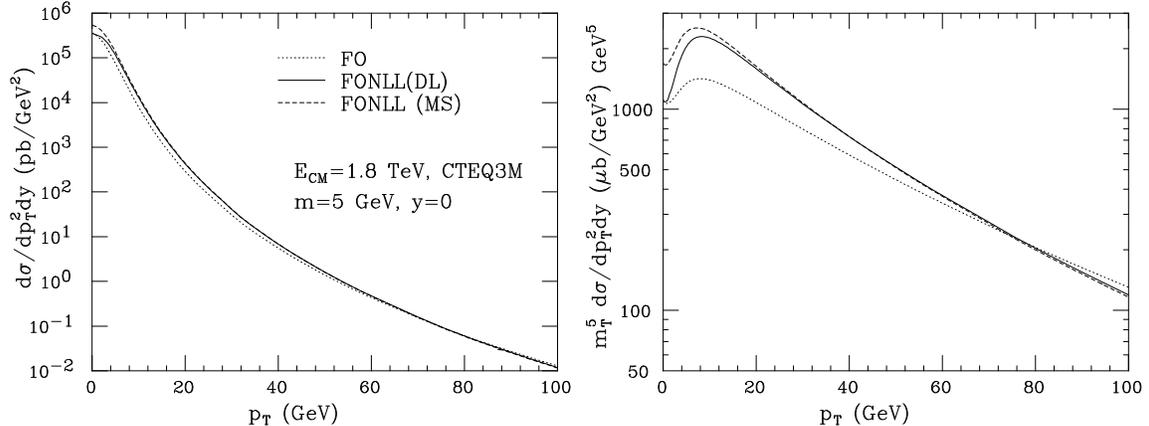

\begin{center}
\epsfig{file=comb.eps,height=5.7cm}
\epsfig{file=comb-mt.eps,height=5.55cm}
\parbox{\capwidth}{
\caption{\label{fig:comb} \protect\small Combined cross section, 
both in the \MSB\ and in the DL schemes, obtained with $c=0$.}
}
\end{center}
%\label{fig:comb}
\end{figure}

We begin by combining the various terms with the choice of scales
$\mur=\muf=\mu=\mt$.  For the sake of discussion, we now use $G(m,\pt)=1$,
corresponding to the choice $c=0$ in eq.~(\ref{eq:merge2}).
Figure~\ref{fig:comb} shows the
full massive result (FO) compared with the combined ones, both in the
\MSB\ and in the DL schemes.

We can first of all appreciate the effect of switching from the \MSB\ to
the DL scheme. In the latter the FONLL result
coincides with the FO one at the $\pt=0$ (and hence
$\mu=\mt=m$) point, as desired, while the result in the \MSB\ scheme
presents an offset. This is directly related to the offset previously
noticed in fig.~\ref{fig:dl}, at the $m=\pt$ point.  It is
worth mentioning how the cancellation in the DL scheme
still relies on our choice of
factorization scale $\muf=\sqrt{m^2+\pt^2}$, which ensures $\muf=m$ at
$\pt=0$. Any other scale choice will destroy the matching at this point,
giving results qualitatively similar to those obtained in the \MSB\ scheme.
This will appear evident when we will present the bands obtained by
varying the renormalization/factorization scales.

\begin{figure}[t]
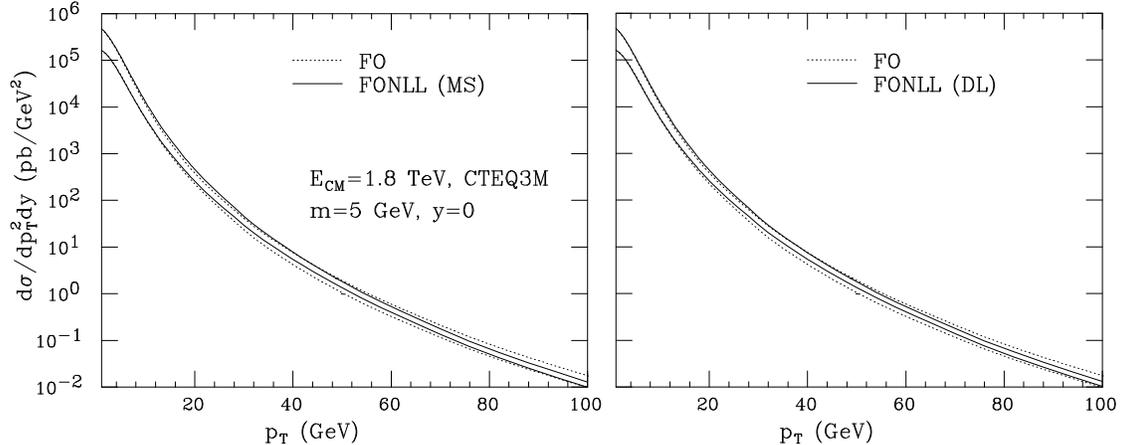

\begin{center}
\epsfig{file=band-ms.eps,height=5.95cm}
\epsfig{file=band-dl.eps,height=5.8cm}
\parbox{\capwidth}{
\caption{\label{fig:bands} \protect\small Combined cross section, 
both in the \MSB\ and in the DL schemes. Shown are the bands obtained
by varying independently the renormalization and factorization scales.
A suppression factor of the small $\pt$ region,  according to eq.~(\protect\ref{eq:merge2}), with $c=5$, is included.}}
\end{center}
%\label{fig:bands}
\end{figure}
\begin{figure}[t]
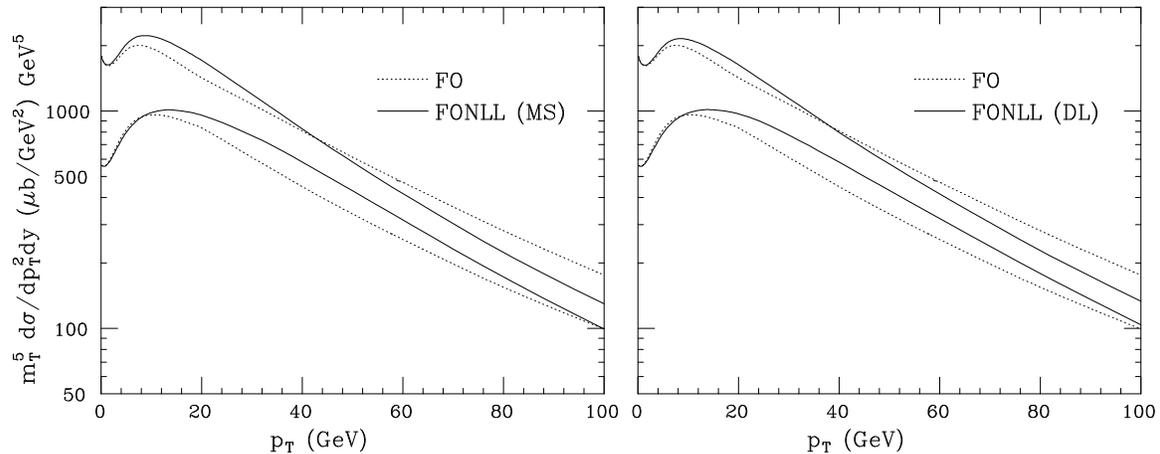

\begin{center}
\epsfig{file=band-ms-mt.eps,height=6cm}
\epsfig{file=band-dl-mt.eps,height=6cm}
\parbox{\capwidth}{
\caption{\label{fig:bands-mt5} \protect\small
Same as in fig.~\ref{fig:bands}, with an $\mt^5$ weight.
}}
\end{center}
%\label{fig:bands}
\end{figure}

We now describe the procedure we follow in order to obtain
our final theoretical result, including the errors estimated
by varying the factorization and renormalization scale.
First of all, we fix $c=5$, as discussed in sect.~\ref{sec:powereff}.
Our prediction band is the envelope of the curves
obtained by varying the renormalization and factorization scales
in the range $[\mt/2,2\mt]$. We also
distinguish in our calculation the factorization scale
for the structure functions from the one for the fragmentation functions,
varying them independently.
The resulting bands, for both the \MSB\ and the DL
schemes, are shown in figs.~\ref{fig:bands} and \ref{fig:bands-mt5}.

A few interesting observations can be made about these plots. First of
all, the band of the combined result is much narrower in the
large-$\pt$ region.  This result had already been found with the resummed
calculation of ref. \cite{Cacciari94}, and is the expected outcome of the
resummation of the large $\log(\pt/m)$ terms. Further, we notice that the
combined bands match exactly the FO ones at $\pt\,\to\,0$. This feature is
due to the suppression factor $G(m,\pt)$, which annihilates
completely the RS$\,-\,$FOM0
term at this point. Without this suppression the band would be
very large in this region, owing to
the uncontrolled behaviour of the massless approximation at small
$\pt$. Last, the combined band lies in the intermediate-$\pt$ region, a
little above the FO one, and is about equally wide.  The maximum
increase is of the order of 10\%--20\% on the NLO full massive result.
We also notice that the DL and \MSB\ schemes give about the same result:
no significant differences can be observed.

\begin{figure}[!tb]
\begin{center}
\epsfig{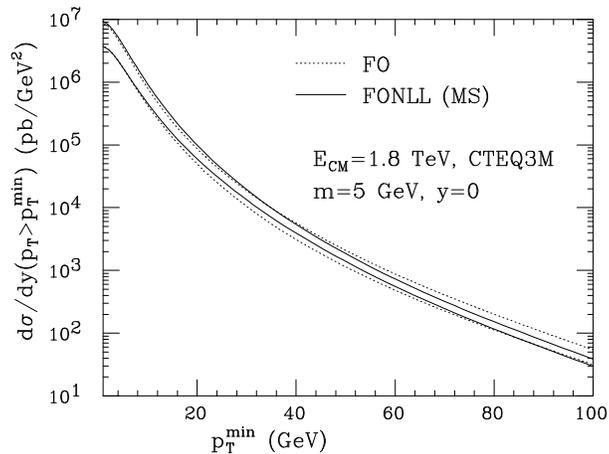}
\parbox{\capwidth}{
\caption{\label{fig:intband} \protect\small
The $b$ cross section per unit rapidity as a function of the $\pt$
cut.}}
\end{center}
%\label{fig:bands}
\end{figure}

The Tevatron results for $b$ production are usually given in terms
of a cross section with a cut
$\pt>\pt^{\rm min}$. In order to give a rough
estimate of the effect of resummation for this quantity, relative to the
NLO calculation, we plot it in fig.~\ref{fig:intband}.
Since this is only included as an indication,
and since the cross section is reasonably uniform in rapidity
in the central region, we do not perform
the $y$ integration in the range $\abs{y}<0.5$.

In summary, we observe that our resummation procedure indicates
the presence of a small enhancement in the intermediate $\pt$ region,
followed by a reduction of the cross section (and of the uncertainty band)
at larger $\pt$.
The question we should now ask is how
reliable this enhancement is.
Of course, if we used smaller values for the
parameter $c$, the cross section at moderate $\pt$ would increase,
thus helping in explaining the discrepancy between theory and the Tevatron
data. However, the considerations given in sect.~\ref{sec:powereff}
do indicate that $c=5$ is the conservative choice we should make
to obtain a reliable result.

\section{Conclusions}
\label{sec:conclusions}
In summary, we have defined a procedure for improving in a systematic
way the calculation of the transverse-momentum heavy-flavour spectrum.
Using already available calculations, this procedure has been
applied to obtain the $\pt$ spectrum of heavy-flavour production
with NLO accuracy in $\as$, and with NLL resummation of
logarithms of $\pt/m$.

This calculation confirms essentially the results of refs.~\cite{Nason89}
and \cite{Cacciari94}, to which it reduces in the small- and
large-$\pt$ limit, respectively.
In the intermediate transverse momentum region, when $\pt$ is roughly
between two times and five times the heavy quark mass, we find a slight
enhancement of the cross section with respect to previous results.

We can understand this enhancement in two ways. In relation
to the calculation of ref.~\cite{Cacciari94}, we find that
mass corrections at order $\as^2$ and $\as^3$, not included there,
are large and positive in this region. In relation to the calculation
of ref.~\cite{Nason89}, instead, we find large resummation corrections,
arising at the next-to-leading order level.
For example, large contributions arise from the notoriously large
corrections to the parton cross sections for the $gg\to g +X$ process,
convoluted with the gluon fragmentation into heavy quark.
These contributions are of order $\as^4$ (if one accounts for the fact that
the gluon fragmentation into a heavy quark is of order $\as$),
and therefore are not included in the $\as^3$
calculation of ref.~\cite{Nason89}.

In our calculation, we had the freedom to suppress resummation
effects of order higher than $\as^3$ in the small-transverse-momentum region.
We included such a suppression, based upon the comparison
of the full massive- and massless-limit calculations at order $\as^3$.
Since we found that the massless limit result is not a good approximation
to the massive one for $\pt\lesssim 5m$, our suppression factor
begins to act precisely in that region.

In the present work, we have only developed the basic formalism
for a reliable computation of the heavy-quark $\pt$ spectrum.
The enhancement we found goes in the direction of favouring
a better comparison with the Tevatron data on $b$ production.
Other ingredients should, however, be included in a full phenomenological
analysis. A non-perturbative fragmentation-function contribution,
for example, should be extracted from $e^+e^-$ data
(see for example \cite{Cacciari97,Colangelo92}) and
carefully included in the computation. Similarly,
theoretical uncertainties arising from the error in the determination
of the strong coupling constant, of the structure functions and
of the $b$-quark mass should be fully assessed in order to give a reliable
theoretical prediction. We will address these problems in future work.

\noindent {\bf Acknowledgements.} \\
One of us (M.C.) would like to thank the Theory Division
at CERN for its kind hospitality during the completion of this work.

This work was supported in part by the EU Fourth Framework
Programme ``Training and Mobility of Researchers'', Network
``Quantum Chromodynamics and the Deep Structure of
Elementary Particles'', contract FMRX-CT98-0194 (DG 12 - MIHT).

\appendix

\section{Perturbative expansion of the heavy-quark parton density}
\label{app:Fexpan}
We now prove the existence of expansions of the form of eqs.~(\ref{eq:Dseries})
and (\ref{eq:Fseries}).
We consider the Mellin transforms of the fragmentation functions
and structure functions.
We will use the notation
\beq
\label{eq:Mellindef}
D_i(\mu_0) = \int \frac{dx}{x} \, x^N \,D_i(x,\mu_0)\;,
\eeq
and a similar one for the structure functions.
Thus, every time that the $x$ dependence is not given explicitly
we refer to the Mellin transforms of the corresponding quantity.

It is well-known that fragmentation functions at a scale $\mu$
are given in terms of the fragmentation functions at a scale $\mu_0$
(in matrix notation)
by an equation of the form
\begin{equation}
  D(\mu) = \left(1+M(\as(\mu),\log \mu/\mu_0)\right) D(\mu_0)\,,
\end{equation}
where $1$ stands for the identity matrix, and $M$ is a matrix whose
elements are power expansions in $\as(\mu)$ and $\log \mu/\mu_0$,
starting at order $\as(\mu)$, and having in all their terms no more powers of
$\log  \mu/\mu_0$ than powers of $\as$. We choose $\mu_0=m$,
replace $D(m)$ with its power expansion in terms of $\as(m)$,
which contains no logarithms,
and replace $\as(m)$ with its power expansion in terms of $\as(\mu)$,
which does contain logarithms, but always with a smaller power than the
corresponding power of $\as(\mu)$. With these operations
we put $D(\mu)$ in the form of eq.~(\ref{eq:Dseries}).

For eq.~(\ref{eq:Fseries}), we begin with a similar equation
for structure functions
\begin{equation}
  F(\mu_0) = \left(1+M(\as(\mu),\log \mu/\mu_0)\right) F(\mu)\;,
\end{equation}
choose $\mu_0=m$, and separate the $h$ and $\bar{h}$ and the light components
$l$
\begin{eqnarray}
  F_{h/\bar{h}}(m) &=& F_{h/\bar{h}}(\mu)
+M_{h/\bar{h},j}(\as(\mu),\log \mu/m) F_j(\mu)\;,\nonumber \\
  F_l(m) &=& F_l(\mu)
+M_{l,j}(\as(\mu),\log \mu/m) F_j(\mu)\;.
\end{eqnarray}
Observe that the index $j$ runs also on $h$ and $\bar{h}$.
The left-hand side of the first two equations can be given as a power expansion
in $\as(m)$, with coefficients proportional to the $F_l(m)$.
They in fact start at order $\as^2$ in the \MSB\ scheme.
The $F_l(m)$ can be obtained from the remaining equations. Furthermore,
$\as(m)$ can be expanded in powers of $\as(\mu)$ as before.
The first two equations can then be solved for
$F_{h/\bar{h}}(\mu)$
(since the coefficients in front of the $F_{h/\bar{h}}(\mu)$ terms
begin with 1, and can therefore be inverted).
The expressions for the $F_{h/\bar{h}}(\mu)$ components
will be linear functions of the $F_l(\mu)$,
with coefficients that are power expansions in $\as(\mu)$ and in $\log \mu/m$,
with no more powers of logarithms than powers of $\as(\mu)$,
as in eq.~(\ref{eq:Fseries}).

\section{The DL scheme}
\label{app:DL}
We now discuss the change of scheme from the \MSB\ to the DL scheme.
This will be given, for the fragmentation functions and for the
next-to-leading evolution kernels, in terms of Mellin
transforms, defined according to eq.~(\ref{eq:Mellindef}).

Using the notation of \cite{FurmanskiPetronzio}, we can describe the evolution
in terms of the variable
\begin{equation}
t=\frac{2}{b_0}\log\frac{\alpha(\mu_0^2)}{\alpha(\mu)}
\end{equation}
instead of $\mu$.
Calling $E(\mu,\mu_0)$ the evolution matrix from $\mu_0$ to $\mu$
(in $N$ space), we have
\begin{eqnarray}
  \frac{dE(\mu,\mu_0)}{dt}&=&
        \left(P_0+\frac{\alpha(\mu)}{2\pi}R\right)E(\mu,\mu_0)
\nonumber \\
  R&=& P_1-\frac{b_1}{2 b_0} P_0\;.
\end{eqnarray}
The fragmentation functions in the \MSB scheme are given at the low scale by
eqs. (\ref{DQQ})--(\ref{DqQ}), which we rewrite here as
\begin{eqnarray}
  D_h(\mu_0) &=& 1
            +\frac{\alpha(\mu_0)}{2\pi}P_{qq}\log\frac{\mu_0^2}{m^2}
                      +\frac{\alpha(\mu_0)}{2\pi}d_1+{\cal O}(\alpha^2)
\nonumber \\
  D_g(\mu_0) &=& \frac{\alpha(\mu_0)}{2\pi}P_{gq}\log\frac{\mu_0^2}{m^2}
+{\cal O}(\alpha^2)
\nonumber \\
  D_{i\ne h,g}(\mu_0) &=& {\cal O}(\alpha^2)\;.
\label{eq:initMS}
\end{eqnarray}
where
\begin{eqnarray}
  d_1 &=& -\int \frac{dx}{x} \, x^N \,
        \Cf \left[\frac{1+x^2}{1-x}(2\log(1-x)+1)\right]_+\nonumber \\
  P_{qq} &=& \int \frac{dx}{x} \, x^N \,
        \Cf \left[\frac{1+x^2}{1-x}\right]_+ \nonumber \\
  P_{qg} &=& \int \frac{dx}{x} \, x^N \,
        \Tf \left[(1-x)^2+x^2\right] \;.
\end{eqnarray}
We want a scheme $\tilde{D}$ where
\begin{equation}
  \tilde{D}_h(\mu_0)=1+\frac{\alpha(\mu_0)}{2\pi}P_{qq}\log\frac{\mu_0^2}{m^2}
\end{equation}
and the other components of the fragmentation function remain identical.

The transformations
\begin{eqnarray}
  \tilde{D}_h(\mu)&=&\left(1-\frac{\alpha(\mu)}{2\pi}d_1\right)
       D_h(\mu)
\nonumber \\
  \tilde{D}_i(\mu) &=& D_i(\mu)\quad\mbox{for $i\ne h$}
\end{eqnarray}
have precisely this property, that is to say they translate
eqs.~(\ref{eq:initMS}) into
\begin{eqnarray}
  \tilde{D}_h(\mu_0) &=& 1+
     \frac{\alpha(\mu_0)}{2\pi}P_{qq}\log\frac{\mu_0^2}{m^2}+{\cal O}(\alpha^2)
\nonumber \\
  \tilde{D}_g(\mu_0) &=&
 \frac{\alpha(\mu_0)}{2\pi}P_{gq}\log\frac{\mu_0^2}{m^2}+{\cal O}(\alpha^2)
\nonumber \\
  \tilde{D}_{i\ne h,g}(\mu_0) &=& {\cal O}(\alpha^2)\;.
\end{eqnarray}

Having changed the initial conditions, we
now need to find the modifications to the evolution equations
and to the short-distance
cross sections.
Defining the matrix
\begin{equation}
  C_{ij}=-\delta_{ij} d_1\quad \mbox{for $i\ne g$,}\quad C_{ig}=C_{gi}=0\,,
\quad C_{gg}=0\;,
\end{equation}
we have
\begin{equation}
  \tilde{D}=\left(1+\frac{\as}{2\pi}C\right) D
\end{equation}
where we imply vector and matrix notation when we omit the indeces.

Since physical quantities should remain unchanged, we must have
\begin{equation}
  \hat{\sigma}_i(\mu) E_{ij}(\mu,\mu_0) D_j(\mu_0)= 
  \tilde{\sigma}_i(\mu) \tilde{E}_{ij}(\mu,\mu_0) \tilde{D}_j(\mu_0)\,.
\end{equation}
To satisfy the above equality, we define
\begin{eqnarray}
  \tilde{E}(\mu,\mu_0)&=&\left(1+\frac{\alpha(\mu)}{2\pi}C \right) E(\mu,\mu_0)
                         \left(1-\frac{\alpha(\mu_0)}{2\pi}C \right)
\nonumber \\
  \tilde{\sigma}(\mu)&=&\hat{\sigma}
              \left(1-\frac{\alpha(\mu)}{2\pi}C \right)
\end{eqnarray}
We now derive the evolution equation for $\tilde{E}$. We use the obvious
observation
that $(1-\alpha(\mu)/2\pi\,C ) \tilde{E}$ obeys the same evolution
equation as $E$
\begin{equation}
  \frac{d}{dt}\left[\left(1-\frac{\alpha(\mu)}{2\pi}C \right)
     \tilde{E}(\mu,\mu_0)\right] =
    \left(P_0+\frac{\alpha(\mu)}{2\pi}R\right)
  \left[\left(1-\frac{\alpha(\mu)}{2\pi}C \right)\tilde{E}(\mu,\mu_0)\right]\;.
\end{equation}
Since
\begin{equation}
  \frac{d\alpha(\mu)}{dt}=-\frac{b_0}{2}\alpha(\mu)
\end{equation}
we get
\begin{eqnarray}
&&  \frac{b_0}{2}\frac{\alpha(\mu)}{2\pi}C\tilde{E}(\mu,\mu_0)
+\left(1-\frac{\alpha(\mu)}{2\pi}C \right)\frac{d\tilde{E}(\mu,\mu_0)}{dt}
=\phantom{aaaaaaaaaa} \nonumber \\&& 
 \phantom{aaaaaaaaaaaaaaaa}
  \left(P_0+\frac{\alpha(\mu)}{2\pi}R\right)
     \left[\left(1-\frac{\alpha(\mu)}{2\pi}C \right)\tilde{E}\right]\,.
\end{eqnarray}
Multiplying both sides on the left by $(1+\alpha(\mu)/2\pi\,C )$,
and neglecting higher-order terms consistently, we get
\begin{equation}
  \frac{d\tilde{E}(\mu,\mu_0)}{dt}=\left(P_0+\frac{\alpha(\mu)}{2\pi}
\left[R+CP_0-P_0C-\frac{b_0}{2}C\right]\right)\tilde{E}(\mu,\mu_0)\;.
\end{equation}
As a last point, we need to specialize our result to the standard
cases in which one separates singlet and non-singlet component
in the evolution equation. It is clear that for the non-singlet components
we have $C_{\ssrm NS}=-d_1$, and
the commutator term is zero. For the gluon and singlet components, the
$C_{\ssrm S}$
matrix has the form
\begin{equation}
  C_{\ssrm S}=
 \begin{array}[c]{|c c|} C_{gg}&C_{gq} \cr C_{qg} & C_{qq}  \end{array}=
  \begin{array}[c]{|c c|} 0&0 \cr 0& -d_1  \end{array} \;.
\end{equation}

Next we describe how to implement the scheme change in the short- distance
partonic kernel cross sections, written in the notation of~\cite{Aversa89}.
The Born cross section has the form (see also eq.~(\ref{resolved}))
\begin{equation} \label{sigmab}
\sigma_b=  \int dx_3\int_{VW/x_3}^{1-(1-V)/x_3}  dv\,
  F_0(v,x_3)\,{\cal L}(x_1,x_2) D(x_3)/x_3^2\,,
\end{equation}
where
\begin{equation}
  x_1=\frac{VW}{v x_3}\,,\quad\quad x_2=\frac{1-V}{(1-v)x_3}\;.
\end{equation}
Observe that the limits on $v$ imply a limit on $x_3$
\begin{equation}
  \frac{VW}{x_3} < 1-\frac{1-V}{x_3}\;\to\; x_3 > 1-V+ VW\;.
\end{equation}
We now introduce a scheme transformation for the fragmentation
function
\begin{equation} \label{schch}
  D(x_3)=\tilde{D}(x_3)+\int dy\,dz\,\delta(x_3-yz)\,p(y)\tilde{D}(z)\;.
\end{equation}
The second term of eq.~(\ref{schch}), combined with the Born term in
eq.~(\ref{sigmab}), gives rise to a correction to the next-to-leading
contribution.
We wish to cast this correction in the same form as the next-to-leading term
in eq.~(\ref{resolved}), so as to implement it easily in the numerical code.

The following change of variables
\newcommand\ww{w^\prime}
\newcommand\vv{v^\prime}
\begin{equation}
  \ww=\frac{vy}{vy+(1-y)}\,,\quad\quad \vv=1-y(1-v)\;,
\end{equation}
with the inverse
\begin{equation}\label{vpwpvy}
  y=1-\vv(1-\ww)\,,\quad\quad v=\frac{\vv\ww}{\vv\ww+1-\vv}\;,
\end{equation}
maps the unit square onto the unit square, thus the integration range
$0<y<1,0<v<1$ is equivalent to the range $0<\ww<1,0<\vv<1$.
The determinant is
\begin{equation}
  dv\,dy=d\vv\,d\ww\,\frac{\vv}{\vv\ww+1-\vv}
\end{equation}
and
\begin{equation}
  x_1=\frac{VW}{\vv\ww z}\,,\quad\quad x_2=\frac{1-V}{(1-\vv)z}\,.
\end{equation}
The limits on $v$ imply
\begin{eqnarray}
  1-V+VW<&z&<1\,,\nonumber \\
 \frac{VW}{z}<&\vv&<1-\frac{1-V}{z} \nonumber \\
 \frac{VW}{z \vv}<&\ww&<1\;.
\end{eqnarray}
In terms of the new variables the cross section becomes
\begin{equation}
\sigma_b =   \int_{1-V+VW}^1 dx_3\int_{VW/x_3}^{1-(1-V)/x_3}  dv
  F_0(v,x_3){\cal L}(x_1,x_2) \tilde{D}(x_3)/x_3^2\;+\; \delta\sigma\;,
\end{equation}
where
\begin{eqnarray}
&&\delta\sigma=
\int_{1-V+VW}^1 dz \int_{VW/z}^{1-(1-V)/z} d\vv \int_{VW/(z\vv)}^1 d\ww
\phantom{aaaaaaaaaaaa}\nonumber \\
&&\phantom{aaaaaaaaaaa}
\frac{\vv}{(\vv\ww+1-\vv)^3}\;F_0(v,yz)\;{\cal L}(x_1,x_2)\;p(y)\;
\tilde{D}(z)/z^2\;.
\end{eqnarray}
The structure of this equation coincides with that of the next-to-leading
term in eq.~(\ref{resolved}), as desired.

Since $p(y)$ is a distribution at $y=1$, it is convenient to rewrite
the above formula as
\begin{eqnarray}
 && \delta\sigma=
\int_{1-V+VW}^1 dz \int_{VW/z}^{1-(1-V)/z} d\vv \Bigg\{
\int_{VW/(z\vv)}^1 d\ww\;p(y)
\nonumber \\
&&
\left[\frac{\vv}{(\vv\ww+1-\vv)^3} F_0(v,yz)\,{\cal L}(x_1,x_2)
-\vv F_0(\vv,z)\,{\cal L}(x^\prime_1,x^\prime_2)\right]\,\tilde{D}(z)/z^2
\nonumber \\
&&+\int_{1-V+VW}^1 dz \int_{VW/z}^{1-(1-V)/z} d\vv
 \vv F_0(\vv,z)\,{\cal L}(x^\prime_1,x^\prime_2)\,\tilde{D}(z)/z^2
\nonumber \\
&&\phantom{aaaaaaaaaaaaaaaaaaaaaaaaaaaaaaaa}
\times \int_{VW/(z\vv)}^1 d\ww\, p(y)\Bigg\}\;,
\end{eqnarray}
where
\begin{equation}
  x_{1/2}^\prime = x_{1/2}|_{\ww=1}\;.
\end{equation}

In the DL scheme
\begin{equation}
 p(y)=\frac{\alpha C_f}{2\pi}\left[
\frac{1+y^2}{1-y}\left(-1-2\log(1-y)\right)\right]_+\,.
\end{equation}
Using the fact that
\begin{equation}
  \int_0^1 p(y)\,dy=0\;,
\end{equation}
we have
\begin{equation}
  \int_{VW/(z\vv)}^1 d\ww\, p(y)=-\int^{VW/(z\vv)}_{-(1-\vv)/\vv}
         p(\vv\ww+1-\vv)\,d\ww=
-\frac{1}{\vv}\int_0^{1-\vv+VW/z} p(y)\,dy\;,
\end{equation}
where
\begin{equation}
 \int_0^y p(y)\,dy=
\frac{\alpha C_f}{2\pi}\left(\log(1-y)(y^2+2y-1)-2y+2\log^2(1-y)\right)\;.
% checked by fortran
%      function p(y)
%      implicit none
%      real * 8 p,y
%      p=(1+y**2)/(1-y)*(-1-2*log(1-y))
%      end
%      implicit none
%      real * 8 p,y,dgauss,res1,res2
%      external p
% 1    write(*,*) ' enter y '
%      read(*,*) y
%      res1=log(1-y)*(y**2+2*y-1)-2*y+2*log(1-y)**2
%      res2=dgauss(p,0.d0,y,1.d-6)
%      write(*,*) res1/res2
%      goto 1
%      end
\end{equation}

%\newpage
%{\bf Acknowledgements}\newline\noindent

\end{document}